\documentclass{epl}

\title{Monte Carlo critical isotherms for Ising lattices}
\author{J.G. Garc\'{\i}a \and J.A. Gonzalo}
\institute{
Departamento de F\'\i sica de Materiales, Universidad Aut\'onoma de Madrid - \\
Cantoblanco, 28049 Madrid, Spain.}


\begin{document}

\maketitle

\begin{abstract}
Monte Carlo investigations of magnetization versus field, $M_c(H)$, at the critical temperature
provide direct accurate results on the critical exponent $\delta^{-1}$ for one, two, three
and four-dimensional lattices: $\delta_{1D}^{-1}$=0, $\delta_{2D}^{-1}$=0.0666(2)$\simeq$1/15,
$\delta_{3D}^{-1}$=0.1997(4)$\simeq$1/5,
$\delta_{4D}^{-1}$=0.332(5)$\simeq$1/3. This type of Monte Carlo data
on $\delta$, which is not easily found in studies of Ising lattices in the current literature, 
as far as we know, defines extremely well the numerical value of this exponent within very 
stringent limits.
\end{abstract}

\section{Introduction}
The spin-1/2 Ising model \cite{Yeomans} has been a remarkably succesful model 
of a short-range interacting system
to study phase transitions in magnetic, order-disorder, ferroelectric systems, etc. In
this model, the spin variable $s_i$ is allowed to take values $\pm$1 on each 
lattice site. The Hamiltonian that rules
the interaction between the spins has the following form

\begin{equation}
\label{IsingHamiltonian}
\mathcal{H}=-J
\sum_{<ij>} s_{i}s_{j}
- H
\sum_{i} s_{i}
\end{equation}

The phase transition of the system results as a consecuence of the cooperative behavior 
of the spins determined by the
first term of this hamiltonian, where $J$ is the exchange energy. The second term contains
the magnetic field $H$, which also influences the ordering of the spins.

The critical exponents of the Ising universality class have been extensively investigated for
a long time by various methods, High (HT) and Low Temperature (LT) expansions, Monte Carlo (MC)
simulations, Field Theoretical (FT) methods, etc. In particular, near the critical 
point $T$$\longrightarrow$$T_c$ and $H$$\longrightarrow$0,
the susceptibility ($\chi$) exponent
($\gamma$ = $\partial$log$\chi^{-1}$/$\partial$log$\mid$$T-T_c$$\mid$),
the correlation length ($\xi$) exponent
($\nu$ = $\partial$log$\xi^{-1}$/$\partial$log$\mid$$T-T_c$$\mid$),
the pair correlation function [$\Gamma(r)$] exponent $\eta$
(where $\Gamma(r)\sim$$1/r^{d-2+\eta}$),
the specific heat exponent
($\alpha$ = $\partial$log$C^{-1}$/$\partial$log$\mid$$T-T_c$$\mid$),
and the spontaneous magnetization exponent
($\beta$ = $\partial$log$M$/$\partial$log$\mid$$T-T_c$$\mid$)
have been repeatedly investigated by MC methods, but no direct MC
investigations of the critical isotherm exponent
($\delta$ = $\partial$log$H$/$\partial$log$M$)
have been published, as far as we know.

\section{Monte Carlo Method}

The determination of  the critical exponent $\delta$ from Monte Carlo data
at the critical isotherm, $M_{c}(H)$ at $T=T_c$, was made in such a way as to
minimize (1) finite size effects, appearing only at $H$ very
close to zero, and (2) saturation effects, appearing at the opposite end
of the sets of data, i.e. at $H$ away from zero, when the
power law $M$$\sim$$H^{1/\delta}$ loses its validity.

Relatively large lattices (of the order of $10^6$ spins or more) with periodic 
boundary conditions and
very closely spaced field intervals ($\Delta$$H$$\sim$0.0005) were used
to approach the critical point.

In order to confirm the value of the critical temperature for each dimensionality,
$M_{s}(T)$ was first determined by means of a Wolff \cite{Wolff} (modified
Swendsen-Wang \cite{Swendsen}) algorithm. However, $M_{c}(H)$ was studied with a standard
Metropolis \cite{Metropolis} algorithm which produces an excellent local and global
thermalization of the whole lattice for one single temperature (in our case
always $T=T_c$). In all cases, 100,000 Monte Carlo steps were
used in order to insure good thermal equilibrium, and 20,000 states were
considered in the partition function at each temperature / field.

It is easy to understand that our Metropolis algorithm, modified for studying
the evolution of magnetization with magnetic field, is sufficient to
investigate $\delta$. When we make calculations modifying $H$ by 
very small amounts one would not expect too large changes 
in the fluctuations, making it unnecessary to
use modified clusters algorithms, which get statistically more independent 
states for the evolution with temperature, as it is well known. In any case, 
the good quality of the results shown
below, obtained with the Metropolis algorithm, speak for themselves.

The $M_{c}(H)$ data allow the direct determination
of $\delta^{-1}$ from log-log plots
of $M$ vs. $H$ at $T=T_c$.

\section{One-dimensional Ising critical isotherm}

The one-dimensional case is the easiest of all to be considered because the
critical temperature is exactly zero (there is no real phase transition). In Figure 1, we
show that the smallest non-zero value of the magnetic field $H$ make the whole lattice turn
in the same direction, perfectly ordered, keeping the magnetization at the maximum
value ($M$=1) all the way.

\begin{figure}
\onefigure[width=6.1cm,height=7.9cm,angle=270]{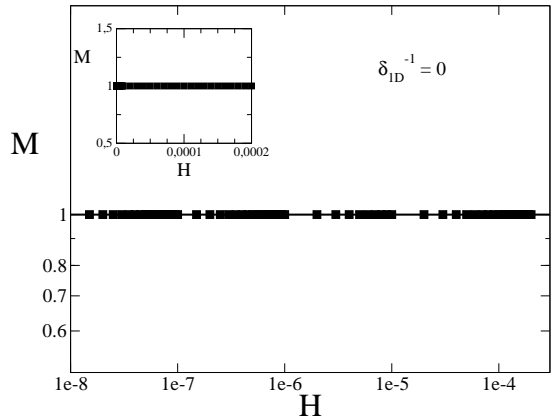}
\caption{Magnetization ($M$) vs. field ($H$) at $T$=$T_c$=0 for a one-dimensional ($d$=1)
Ising chain. The invariant magnetization confirms the
expected $\delta_{1D}^{-1}$=0 value for the critical exponent.}
\label{fig1}
\end{figure}

\section{Two-dimensional Ising critical isotherm}

In 1944, Onsager \cite{Onsager} performed the calculation of the exact partition function of the
two-dimensional Ising model in zero field. However, the same problem in a magnetic field
remain unsolved although we can study their behavior numerically.
In Figure 2, numerical Monte Carlo $M_c(H)$ data allow the direct determination
of $\delta_{2D}^{-1}$=0.0666(2) from log-log plots of $M$ vs. $H$
at $T$=$T_c$=2.269185314213.

\begin{figure}
\onefigure[width=6.1cm,height=7.9cm,angle=270]{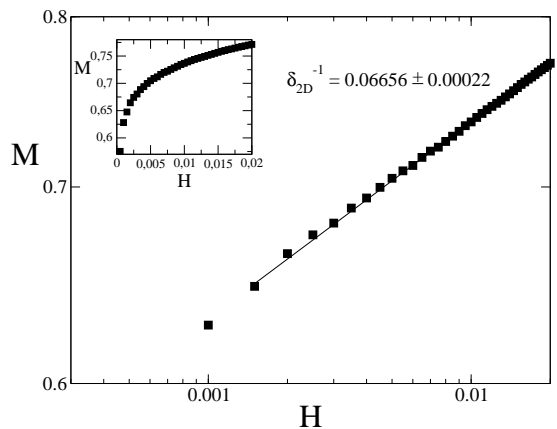}
\caption{Magnetization ($M$) vs. field ($H$) at $T$=$T_c$=2.269185314213 for a two-dimensional
($d$=2) Ising squared monolayer. The slope gives the $\delta_{2D}^{-1}$=0.0666(2) value for the
critical isotherm exponent.}
\label{fig2}
\end{figure}

\section{Three-dimensional Ising critical isotherm}

The Ising criticality for dimension $d$=3 has been studied using several theoretical
approaches (including MC simulations) that have resulted in an (almost complete) set
of safe estimates of the critical exponents. Some exponents like $\gamma$ or $\beta$ have
received special attention and their best estimates have been summarized \cite{Pelissetto}
as $\gamma_{3D}$$\simeq$1.2372(5) and $\beta_{3D}$$\simeq$0.3265(3). However, in the case
of the critical exponent $\delta_{3D}$, there is no direct MC
measure of its value and the currently accepted estimate have been determined using the scaling
relation $\gamma$=$\beta$($\delta$-1) resulting in
a value of $\delta_{3D}$=4.789(2).

As it is shown if Figure 3, our Monte Carlo $M_c(H)$ data allow the determination
of $\delta_{3D}^{-1}$=0.1997(4) from log-log plots of $M$ vs. $H$ at the currently accepted
value \cite{Blote} of the critical temperature $T$=$T_c$=4.511523785. Note that this direct value of the
critical exponent $\delta_{3D}$$\simeq$5 is appreciably different from the actual \cite{Pelissetto}
estimated value 4.789(2).

\begin{figure}
\onefigure[width=6.1cm,height=7.9cm,angle=270]{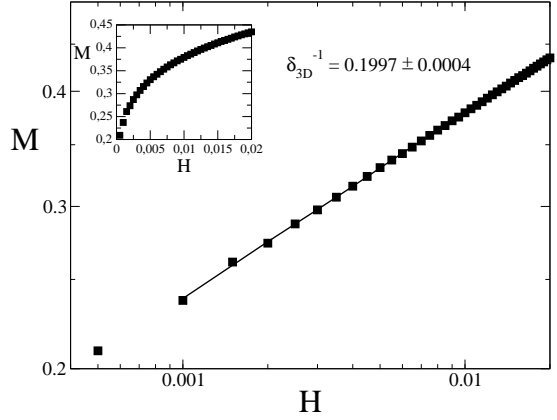}
\caption{Magnetization ($M$) vs. field ($H$) at $T$=$T_c$=4.511523785 for a three-dimensional
($d$=3) Ising cube. The slope gives the $\delta_{3D}^{-1}$=0.1997(4) value for the
critical exponent.}
\label{fig3}
\end{figure}

\section{Four-dimensional Ising critical isotherm}

It is important to note at this point, that for dimensions $d$$\geq$4 the exponents of
the Ising model become the same and take the corresponding mean-field 
values \cite{Gonzalo}. These exponents suddenly lock
into a set of values that become independent of the dimensionality.

In Figure 4, our Monte Carlo $M_c(H)$ data allow the determination
of $\delta_{4D}^{-1}$=0.332(5) from log-log plots of $M$ vs. $H$ at the currently accepted
value \cite{Stauffer} of the critical temperature $T$=$T_c$=6.68029(9). It must be noted that
finite size effects begin to be observable just at $H$$\leq$0.001 for the particular 
lattice size used.

\begin{figure}
\onefigure[width=6.1cm,height=7.9cm,angle=270]{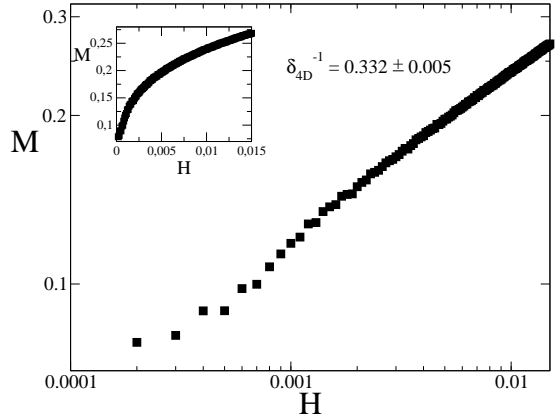}
\caption{Magnetization ($M$) vs. field ($H$) at $T$=$T_c$=6.68029 for a four-dimensional
($d$=4) Ising hyper-cube. The slope gives the $\delta_{4D}^{-1}$=0.332(5) value for the
critical exponent.}
\label{fig4}
\end{figure}

\section{Concluding remarks}

Table I summarizes our results for Ising systems (1$\leq$$d$$\leq$4) showing excellent
agreement between our MC results for $\delta$ and the respective integer values $\delta$($d$)
indicated.

\begin{table}
\caption{Numerical values for transition temperature ($T_c$) and critical isotherm
exponent ($\delta_{MC}^{-1}$) for Ising systems with 1$\leq$$d$$\leq$4.
* Note that the value for $d$=3 is appreciably different from the currently
accepted value $\delta_{3D}$=4.789(2) \cite{Pelissetto}.}
\label{t.1}
\begin{center}
\begin{tabular}{lcccc}

$d$ & 1 & 2 & 3 & 4 \\
\hline
$T_c$ & 0 & 2.269185314213 & 4.511523785 & 6.68029 \\
$\delta_{MC}^{-1}$ & 0 & 0.0666(2) & 0.1997(4) & 0.332(5) \\
$\delta$($d$) & $\infty$ & 15 & $5^{*}$ & 3 \\

\end{tabular}
\end{center}
\end{table}

\section{Acknowledgments}
We specially acknowledge helpful comments and software by M.I. Marqu\'es.
Support from the Spanish DGICyT through Grant Number BFM2000-0032 is gratefully acknowledged.



\begin{thebibliography}{0}

\bibitem{Yeomans}
  \Name{J.M. Yeomans}
  \Book{Statistical Mechanics of Phase Transitions}
  \Publ{Oxford University Press}
  \Year{1992}.

\bibitem{Wolff}
  \Name{U. Wolff}
  \REVIEW{Phys. Rev. Lett.}{62}{1989}{361}.

\bibitem{Swendsen}
  \Name{R.H. Swendsen \and J.S. Wang}
  \REVIEW{Phys. Rev. Lett.}{58}{1987}{86}.

\bibitem{Metropolis}
  \Name{N. Metropolis, A.W. Rosenbluth, M.N. Rosenbluth, A.H. Teller \and E. Teller}
  \REVIEW{J. Chem. Phys.}{21}{1953}{1087}.

\bibitem{Onsager}
  \Name{L. Onsager}
  \REVIEW{Phys. Rev.}{65}{1944}{117}.

\bibitem{Pelissetto}
  \Name{A. Pelissetto \and E. Vicari}
  \REVIEW{Phys. Rep.}{368}{2002}{549-727}.

\bibitem{Blote}
  \Name{H.W.J. Bl\"ote, L.N. Shchur and A.L. Talapov}
  \REVIEW{Int. J. Mod. Phys. C}{10}{1999}{137}.

\bibitem{Gonzalo}
  \Name{J.A. Gonzalo}
  \Book{Effective Field Approach to Phase Transitions and some Applications to Ferroelectrics}
  \Publ{World Scientific, Singapore}
  \Year{1991}.

\bibitem{Stauffer}
  \Name{D. Stauffer \and J. Adler}
  \REVIEW{Int. J. of Mod. Phys. C, Vol. 8}{2}{1997}{263-267}.

\end{thebibliography}
\end{document}